\documentclass[12pt]{article}
 \usepackage{epsfig}
 \def\be{\begin{equation}}
 \def\ee{\end{equation}}
 \def\bea{\begin{eqnarray}}
 \def\eea{\end{eqnarray}}
 \usepackage{graphicx}

 \catcode`\@=11
 \def\lsim{\mathrel{\mathpalette\@versim<}}
 \def\gsim{\mathrel{\mathpalette\@versim>}}
 \def\@versim#1#2{\vcenter{\offinterlineskip
 \ialign{$\m@th#1\hfil##\hfil$\crcr#2\crcr\sim\crcr } }}
 \catcode`\@=12

 \catcode`@=12
\begin{document}
 \thispagestyle{empty}
 \begin{flushright}
 UCRHEP-T597\\
 Mar 2019\
 \end{flushright}
 \vspace{0.6in}
 \begin{center}
 {\LARGE \bf Scotogenic $U(1)_\chi$ Dirac Neutrinos\\}
 \vspace{1.2in}
 {\bf Ernest Ma\\}
 \vspace{0.2in}

{\sl Physics and Astronomy Department,\\ 
University of California, Riverside, California 92521, USA\\}
\vspace{0.1in}
\end{center}
 \vspace{1.2in}

\begin{abstract}\
The standard model of quarks and leptons is extended to include the gauge 
symmetry $U(1)_\chi$ which comes from $SO(10) \to SU(5) \times U(1)_\chi$. 
The radiative generation of Dirac neutrino masses through dark matter is 
discussed in two examples.  One allows for light Dirac fermion dark matter. 
The other allows for self-interacting scalar dark matter with a light scalar 
mediator which decays only to two neutrinos.
\end{abstract}

 \newpage
 \baselineskip 24pt

\noindent \underline{\it Introduction}~:~
Whereas neutrinos are usually assumed to be Majorana, there is yet no 
experimental evidence, i.e. no definitive measurement of a nonzero 
neutrinoless double beta decay.  To make a case for neutrinos to be Dirac, 
the first is to justify the existence of a right-handed neutrino $\nu_R$, 
which is not necessary in the standard model (SM) of quarks and leptons. 
An obvious choice is to extend the SM gauge symmetry 
$SU(3)_C \times SU(2)_L \times U(1)_Y$ to the left-right symmetry 
$SU(3)_C \times SU(2)_L \times SU(2)_R \times U(1)_{(B-L)/2}$. 
In that case, the $SU(2)_R$ doublet $(\nu,e)_R$ is required, and the 
charged $W_R^\pm$ gauge boson is predicted along with a neutral $Z'$ gauge 
boson.

A more recent choice is to consider $U(1)_\chi$ which comes from 
$SO(10) \to SU(5) \times U(1)_\chi$, with $SU(5)$ breaking to the SM. 
Assuming that $U(1)_\chi$ survives to an intermediate scale, the current 
experimental bound on the mass of $Z_\chi$ being about 
4.1 TeV~\cite{atlas-chi-17,cms-chi-18}, then $\nu_R$ 
must exist for the cancellation of gauge anomalies.  Now $\nu_R$ is a 
singlet and $W_R^\pm$ is not predicted.  In this context, new insights 
into dark matter~\cite{m18, m19-1} and Dirac neutrino masses~\cite{m19-2} 
have emerged.  In particular, it helps with the following second issue 
regarding a Dirac neutrino mass.  Since neutrino masses are known to be 
very small, the corresponding Yukawa couplings linking $\nu_L$ to $\nu_R$ 
through the 
SM Higgs boson must be very small.  To avoid using such a small coupling, 
a Dirac seesaw mechanism~\cite{rs84,m02} is advocated in Ref.~\cite{m19-2}. 
The alternative is to consider radiative mechanisms, especially through 
dark matter, called scotogenic from the Greek 'scotos' meaning darkness. 
Whereas the original idea~\cite{m06} was applied to Majorana neutrinos,  
one-loop~\cite{gs08,fm12} and two-loop~\cite{bmpv16} examples for Dirac 
neutrinos already exist in the context of the SM.  For a generic discussion 
of Dirac neutrinos, see Ref.~\cite{mp17}, which is patterned after that 
for Majorana neutrinos~\cite{m98}. Here two new $U(1)_\chi$ examples 
are shown.  One allows for light Dirac fermion dark matter. The other 
allows for self-interacting scalar dark matter with a light scalar 
mediator which decays only to two neutrinos.

\noindent \underline{\it First Scotogenic $U(1)_\chi$ Model}~:~
The particle content follows that of Ref.~\cite{m19-2} except for the 
addition of $\zeta \sim (1,15)$ from the 672 of $SO(10)$. 
This is used to break $U(1)_\chi$ without breaking global lepton number. 
The fermions are shown in Table 1 and scalars in Table 2. 
\begin{table}[tbh]
\centering
\begin{tabular}{|c|c|c|c|c|c|c|c|c|c|c|}
\hline
fermion & $SO(10)$ & $SU(5)$ & $SU(3)_C$ & $SU(2)_L$ & $U(1)_Y$ & 
$U(1)_\chi$ & $Z_2^A$ & $Z_2^B$ & $Z_2^C$ & $Z_2^D$ \\
\hline
$d^c$ & 16 & $5^*$ & $3^*$ & 1 & 1/3 & 3 & + & + & + & +\\ 
$(\nu,e)$ & 16 & $5^*$ & 1 & 2 & $-1/2$ & 3 & + & + & + & + \\ 
\hline
$(u,d)$ & 16 & 10 & 3 & 2 & 1/6 & $-1$ & + & + & + & + \\ 
$u^c$ & 16 & 10 & $3^*$ & 1 & $-2/3$ & $-1$ & + & + & + & + \\ 
$e^c$ & 16 & 10 & 1 & 1 & 1 & $-1$ & + & + & + & + \\ 
\hline
$\nu^c$ & 16 & 1 & 1 & 1 & 0 & $-5$ & $-$ & $-$ & $-$ & $-$ \\ 
\hline
$N$ & $126^*$ & 1 & 1 & 1 & 0 & 10 & $-$ & + & $-$ & + \\ 
$N^c$ & 126 & 1 & 1 & 1 & 0 & $-10$ & + & + & $-$ & $-$ \\ 
\hline
\end{tabular}
\caption{Fermion content of model.}
\end{table}
\begin{table}[tbh]
\centering
\begin{tabular}{|c|c|c|c|c|c|c|c|c|c|c|}
\hline
scalar & $SO(10)$ & $SU(5)$ & $SU(3)_C$ & $SU(2)_L$ & $U(1)_Y$ & 
$U(1)_\chi$ & $Z_2^A$ & $Z_2^B$ & $Z_2^C$ & $Z_2^D$ \\
\hline
$(\phi_1^0,\phi_1^-)$ & 10 & $5^*$ & $1$ & 2 & $-1/2$ & $-2$ & + & + & + & +\\ 
$(\phi_2^+,\phi_2^0)$ & 10 & $5$ & 1 & 2 & $1/2$ & 2 & + & + & + & + \\ 
\hline
$(\eta^+,\eta^0)$ & 144 & 5 & 1 & 2 & 1/2 & $7$ & + & + & $-$ & $-$ \\ 
$\sigma$ & 16 & 1 & $1$ & 1 & $0$ & $-5$ & + & $-$ & + & $-$ \\ 
\hline
$\zeta$ & 672 & 1 & 1 & 1 & 0 & 15 & + & + & + & + \\
\hline
\end{tabular}
\caption{Scalar content of model.}
\end{table}

New fermions $N,N^c$ belonging to $126^*,126$ 
respectively are added per family, as well as a Higgs doublet from 144 
and a singlet from 16.  Note that their $Q_\chi$ charges are fixed by the 
$SO(10)$ representations from which they come.  It should also be clear that 
incomplete $SO(10)$ and $SU(5)$ multiplets are considered here (which is the 
case for all realistic grand unified models).  
Since $\Phi_1^\dagger$ transforms exactly like $\Phi_2$, the linear 
combination $\Phi = (v_1 \Phi_1^\dagger + v_2 \Phi_2)/\sqrt{v_1^2+v_2^2}$ 
is the analog of the standard-model Higgs doublet, where 
$\langle \phi^0_{1,2} \rangle = v_{1,2}$. 
An important $Z_2$ discrete symmetry is imposed so that 
$\nu^c$ is odd and all other SM fields are even, preventing thus the 
tree-level Yukawa coupling $(\nu \phi^0 - e \phi^+)\nu^c$. 
This $Z_2$ symmetry is respected by all dimension-four terms of the 
Lagrangian.  It will be broken softly by the dimension-three trilinear term 
$\mu \sigma \Phi^\dagger \eta$ (in cases B and D) or the $m_N N N^c$ mass 
term (in cases A and C). This allows the one-loop diagram of Fig.~1 
to generate a radiative Dirac neutrino mass.
\begin{figure}[htb]
\vspace*{-5cm}
\hspace*{-3cm}
\includegraphics[scale=1.0]{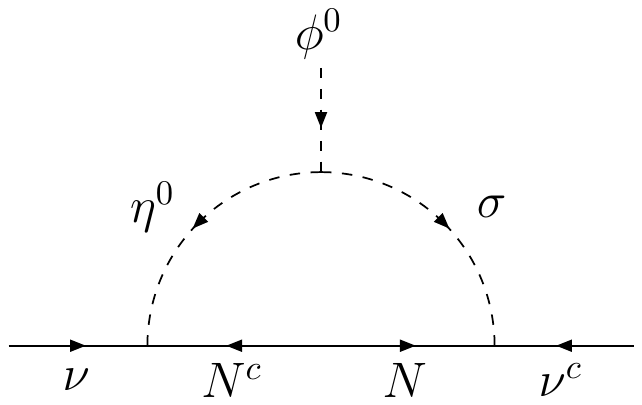}
\vspace*{-21.5cm}
\caption{First one-loop diagram for scotogenic $U(1)_\chi$ Dirac neutrino 
mass.}
\end{figure}
Cases C and D allow the Yukawa coupling $\zeta \nu^c N^c$ which would violate 
lepton number, hence only cases A and B will be considered.  In case A, the 
quartic scalar term $\zeta \sigma^3$ is allowed.  Hence the would-be dark 
U(1) symmetry is reduced to $Z_3$, i.e. $\omega$ for $\sigma,N^c$ and 
$\omega^2$ for $\eta,N$, where $\omega = \exp(2\pi i/3)$.  In case B, 
it is forbidden, so the model possesses a dark U(1) symmetry, i.e. 
$D$ is 1 for $\sigma,N^c$ and $-1$ for $\eta,N$.  In either case, there is 
still a conserved lepton symmetry, i.e. $L=1$ for $\nu,N$ and $L=-1$ for 
$\nu^c,N^c$.  The idea of using a scalar which breaks a gauge U(1) symmetry 
by 3 units, so that a global U(1) symmetry remains was first discussed 
in Ref.~\cite{mpr13} and then used for $B-L$ in Ref.~\cite{ms15}. 
There have been also studies~\cite{yd18,bccps18,cryz18}, using 
dimension-five operators, i.e. $(\nu \phi^0 - e \phi^+)\nu^c S/\Lambda$ 
where $\nu^c$ carries a new charge which forbids the dimension-four 
term but the singlet scalar $S$ carries a compensating charge which allows the 
dimension-five term.

To compute the neutrino mass of Fig.~1, note first that it is equivalent 
to the difference of the exchanges of two scalar mass eigenstates
\begin{equation}
\chi_1 = \cos \theta ~\sigma - \sin \theta ~\bar{\eta}^0, ~~~ 
\chi_2 = \sin \theta ~\sigma + \cos \theta ~\bar{\eta}^0,  
\end{equation}
where $\theta$ is the mixing angle due to the $\bar{\phi}^0 \eta^0 \sigma$ 
term.  Let the $\nu_i N^c_k \eta^0$ Yukawa coupling be $h^L_{ik}$ and 
the $\nu^c_j N_k \sigma$ Yukawa coupling be $h^R_{jk}$, then the 
Dirac neutrino mass matrix is given by
\begin{equation}
({\cal M_\nu})_{ij} = \sum_k {h^L_{ik} h^R_{jk} \sin 2 \theta M_k \over 16 
\pi^2} \left[ {m_2^2 \over m_2^2-M_k^2} \ln {m_2^2 \over M_k^2} - 
{m_1^2 \over m_1^2-M_k^2} \ln {m_1^2 \over M_k^2} \right],
\end{equation}
where $m_{1,2}$ are the masses of $\chi_{1,2}$ and $M_k$ is the mass of 
$N_k$.  If $|m_2^2-m_1^2| << m_2^2 + m_1^2 = 2m_0^2 << M_k^2$, then 
\begin{equation}
({\cal M_\nu})_{ij} = \sum_k {\sin 2 \theta (m_2^2-m_1^2) \over 16 \pi^2 M_k} 
h^L_{ik} h^R_{jk} \left[ \ln {M_k^2 \over m_0^2} - 1 \right].
\end{equation}
This expression is of the radiative seesaw form.  On the other hand, 
if $M_k << m_{1,2}$, then~\cite{m12}
\begin{equation}
({\cal M_\nu})_{ij} =  {\sin 2 \theta \ln(m_2^2/m_1^2) \over 16 \pi^2} 
\sum_k h^L_{ik} h^R_{jk} M_k.
\end{equation}
This is no longer a seesaw formula.  It shows that the three Dirac neutrinos 
$\nu$ have masses which are linear functions of the three light dark Dirac 
fermions $N$.  This interesting possibility opens up the parameter space 
in the search for fermion dark matter with masses less than a few GeV.
 
Consider the annihilation of $N \bar{N} \to \nu \bar{\nu}$ 
through $\chi_1$ exchange, assuming that $\theta$ is very small in Eq.~(1). 
The cross section $\times$ relative velocity is 
\begin{equation}
\sigma \times v_{rel} = {(h^R)^4 \over 32 \pi^2} {m_N^2 \over 
(m_1^2 + m_N^2)^2}.
\end{equation}
As an example, let $m_N = 6$ GeV, $m_1 = 100$ GeV, $h^R = 0.92$, then 
this is about 1~pb, which is the correct value for $N$ to have the 
observed dark-matter relic abundance of the Universe, i.e. 
$\Omega h^2 = 0.12$. 
In Eq.~(4), let $\sin 2 \theta = 10^{-4}$, $h^L =10^{-4}$, and $m_2 = 115$ 
GeV, then $m_\nu = 0.1$ eV, as desired.

At the mass of 6 GeV, the constraint on the elastic scattering cross 
section of $N$ off nuclei is about $2.5 \times 10^{-44}$~cm$^2$ from 
the latest XENON result~\cite{xenon18}.  This puts a lower limit on the 
mass of $Z_\chi$, i.e. 
\begin{equation}
\sigma_0 = {4m_P^2 \over \pi} {[Z f_P +(A-Z) f_N]^2 \over A^2} 
< 2.5 \times 10^{-8}~{\rm pb},
\end{equation}
where
\begin{equation}
f_P = g^2_{Z_\chi} N_V (2u_V + d_V)/M^2_{Z_\chi}, ~~~ 
f_N = g^2_{Z_\chi} N_V (u_V + 2d_V)/M^2_{Z_\chi}, ~~~ 
\end{equation}
and $Z=54$, $A=131$ for xenon.  In $U(1)_\chi$, the vector couplings are 
\begin{equation}
N_V = \sqrt{5 \over 2}, ~~~ u_V =0, ~~~ d_V = {-1 \over \sqrt{10}}.
\end{equation}
Using $\alpha_\chi = g^2_{Z_\chi}/4\pi = 0.0154$ from Ref.~\cite{m18}, 
the bound $M_{Z_\chi} > 4.5$ TeV is obtained.

\noindent \underline{\it Second Scotogenic $U(1)_\chi$ Model}~:~
Using two new fermion singlets and one fermion doublet, 
with a different $Z_2$, another one-loop diagram is obtained in Fig.~2.
\begin{figure}[htb]
\vspace*{-5cm}
\hspace*{-3cm}
\includegraphics[scale=1.0]{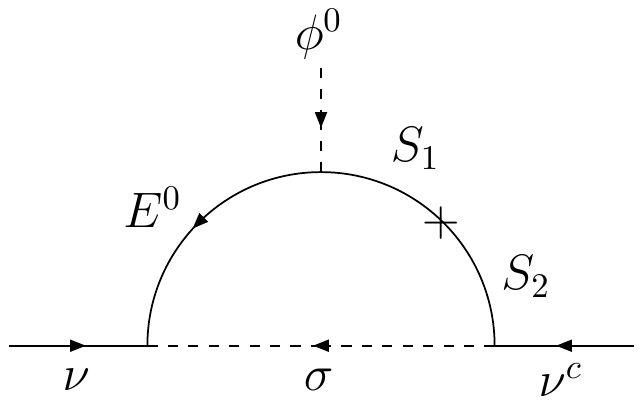}
\vspace*{-21.5cm}
\caption{Second one-loop diagram for scotogenic $U(1)_\chi$ Dirac neutrino 
mass.}
\end{figure}
The relevant particles are shown in Table~3.
\begin{table}[tbh]
\centering
\begin{tabular}{|c|c|c|c|c|c|c|c|c|c|c|c|c|}
\hline
particle & $SO(10)$ & $SU(5)$ & $SU(3)_C$ & $SU(2)_L$ & $U(1)_Y$ & 
$U(1)_\chi$ & $Z_2$ & $Z_4^L$ & $Z_2^{(D)}$ \\
\hline
$(\nu,e)$ & 16 & $5^*$ & 1 & 2 & $-1/2$ & 3 & + & $i$ & + \\ 
$\nu^c$ & 16 & 1 & 1 & 1 & 0 & $-5$ & $-$ & $-i$ & + \\ 
\hline
$(E^+,E^0)$ & 10 & 5 & 1 & 2 & 1/2 & 2 & $-$ & 1 & $-$ \\ 
$S_1$ & 45 & 24 & 1 & 1 & 0 & 0 & $-$ & 1 & $-$ \\ 
$S_2$ & 45 & 24 & 1 & 1 & 0 & 0 & $+$ & 1 & $-$ \\ 
\hline
\hline
$\sigma$ & 16 & 1 & $1$ & 1 & $0$ & $-5$ & $-$ & $-i$ & $-$ \\ 
$\zeta'$ & 126 & 1 & 1 & 1 & 0 & $-10$ & + & $-1$ & + \\
\hline
$\zeta''$ & 2772 & 1 & 1 & 1 & 0 & $-20$ & + & 1 & + \\
\hline
\end{tabular}
\caption{Fermion and scalar content of model.}
\end{table}
Again, the $Z_2$ symmetry forbids the would-be tree-level Yukawa coupling 
$\phi^0 \nu \nu^c$, but is softly broken by the $S_1 S_2$ 
mass term, whereas $S_1^2$ and $S_2^2$ are allowed Majorana mass terms. 
The $U(1)_\chi$ gauge symmetry is broken by $\zeta''$ and since it couples 
to $(\zeta')^2$ and $\zeta'$ couples to $\sigma^2$, the residual symmetry 
of this model is $Z_4$~\cite{hr13,h13,cmsv17,csv16,hsv18}, which enforces 
the existence of Dirac neutrinos, and the dark symmetry is $Z_2$, 
i.e. $(-1)^{Q_\chi+2j}$ as pointed out in Refs.~\cite{m18,kkr09}, 
as shown in Table 3.
 
In this second model, the scalar $\sigma$ is a pure singlet, whereas 
in the first model, it must mix with $\eta^0$ which is part of a 
doublet.  Because of 
the $\zeta' \sigma \sigma$ interaction, it is a self-interacting 
dark-matter candidate~\cite{kkpy17} which can explain the flatness of the 
core density profile of dwarf galaxies~\cite{detal09} and other related 
astrophysical phenomena.  The light scalar mediator $\zeta'$ decays 
dominantly to $\nu^c \nu^c$ so it does not disturb~\cite{gibm09} 
the cosmic microwave background (CMB)~\cite{planck16}, thus avoiding the 
severe constraint~\cite{bksw17} due to the enhanced Sommerfeld production 
of $\zeta'$ at late times if it decays to electrons and photons, as in 
most proposed models.  This problem is solved if the light mediator is 
stable~\cite{m17,m18-2,dsw18} or if it decays into $\nu \nu$ through a 
pseudo-Majoron in the singlet-triplet model of neutrino mass~\cite{mm17}. 
A much more natural solution is for it to decay into $\nu^c\nu^c$ as 
first pointed out in the prototype model of Ref.~\cite{m18-1} and 
elaborated in Refs.~\cite{m18,m19-2}.   Here it is shown how it may 
arise in the scotogenic 
Dirac neutrino context using $U(1)_\chi$.  The connection of lepton parity  
to simple models of dark matter was first pointed out in Ref.~\cite{m15}. 
To obtain three 
massive Dirac neutrinos, there are presumably also three $\sigma$'s. 
Only the lightest is stable, the others would decay into the lightest 
plus $\zeta'$ which then decays into two neutrinos.  Typical mass 
ranges for $\sigma$ and $\zeta'$ are
\begin{equation}
100 < m_\sigma < 200~{\rm GeV}, ~~~ 10 < m_{\zeta'} < 100~{\rm MeV}, 
\end{equation}
as shown in Ref.~\cite{m18-1}.  Lastly, the conjugate fermions to 
$(E^+,E^0)$ are also assumed, to allow them to have invariant Dirac masses 
and to cancel the gauge $U(1)_\chi$ anomalies.

\noindent \underline{\it Concluding Remarks}~:~
The $U(1)_\chi$ gauge symmetry and a suitably chosen particle content 
with a softly broken $Z_2$ symmetry are the ingredients for the 
radiative generation of Dirac neutrino masses through dark matter. 
Both the symmetries for maintaing the Dirac nature of neutrinos and 
the stability of dark matter are consequences.  In the first 
example, because the breaking of $U(1)_\chi$ is by 3 units of lepton 
number through the relationship
\begin{equation}
15 (B-L) = 12 Y - 3 Q_\chi,
\end{equation} 
global U(1) lepton number remains, whereas the dark symmetry is either 
$Z_3$ or U(1).  The dark-matter candidate is a Dirac fermion which may 
be light. 
In the second example, the lepton symmetry is $Z_4$ and the dark 
parity is $(-1)^{Q_\chi+2j}$.  The dark-matter candidate is a complex 
scalar which has self-interactions through a light scalar mediator which 
decays only into two neutrinos.  Both cases are interesting variations 
of basic dark matter, and will face further scrutiny in future 
experiments.

\noindent \underline{\it Acknowledgement}~:~
This work was supported in part by the U.~S.~Department of Energy Grant 
No. DE-SC0008541.  I thank Jose Valle and associates (IFIC, Valencia, Spain) 
and Alfredo Aranda and associates (DCPIHEP, Colima, Mexico) for their great 
hospitality during two recent visits which inspired and facilitated this work.

\bibliographystyle{unsrt}

\end{document}